# Preferential Gas Adsorption at the Graphene-Water interface

Hsien-Chen Ko[1], Wei-Hao Hsu[1,2], Chih-Wen Yang[1], Chung-Kai Fang[1], Yi-Hsien Lu[1], Ing-Shouh Hwang[1,2]*

[1]Institute of Physics, Academia Sinica, Nankang, Taipei 115, Taiwan

[2]National Tsing-Hua University, Department of Materials Science and Engineering, Hsinchu 300, Taiwan

Correspondence and requests for materials should be addressed to I.-S.H. (email:ishwang@phys.sinica.edu.tw).

**The contact of water with graphene is of fundamental importance and of great interest for numerous promising applications[1-4], but how graphene interacts with water remains unclear. Here we used atomic force microscopy (AFM) to investigate hydrophilic mica substrates with some regions covered by mechanically exfoliated graphene layers in water. In water containing air gas close to the saturation concentration (within ~40%), cap-shaped nanostructures and ordered stripe domains were observed on graphene-covered regions, but not on pure mica regions. These structures did not appear on graphene when samples were immersed in highly degassed water, indicating that their formation was caused by adsorption of gas dissolved in water. Thus, atomically thin graphene, even at a narrow width of 20 nm, changes the local surface chemistry of a highly hydrophilic substrate. Further, surface hydrophobicity significantly affects gas adsorption, which has broad implications for diverse phenomena in water.**

    Graphene research has become a hot subject due to this material's extraordinary physical and chemical properties as well as the many possible applications of atomically thin graphene layers. In several promising applications, including nanopore sequencing[1], water desalination[2], energy storage[3], and graphene liquid cells[4], graphene contacts water. Thus, direct imaging and

characterization of the graphene-water interface is critical. Several recent studies have focused on the wettability of graphene based on a macroscopic technique, water contact angle (WCA) measurements. An early study suggested complete wetting transparency for monolayer graphene on certain solids, such as Cu, Au, and Si[5], leading the authors to conclude that single-layer graphene has a negligible effect on the interaction between water and solid substrates. In contrast, subsequent studies suggested that single-layer graphene exhibits null or partial wetting transparency[6,7]. Despite this discrepancy, these studies were all supported by WCA measurements, molecular dynamic simulations, and theoretical calculations[8]. A more recent study suggested that earlier WCA experiments on the wettability of graphitic surfaces may have been affected by unintentional hydrocarbon contamination from ambient air[9]. Here we directly imaged the graphene-water interface with AFM in water at nanometer resolution. Our study shows that air gas (mainly nitrogen and oxygen) dissolved in water tends to adsorb to and cover the graphene-water interface, a scenario that was not considered in previous experimental and theoretical investigations.

It is known that gases dissolved in water may accumulate at solid-water interfaces and form soft cap-shaped nanostructures, which are often called surface nanobubbles or interfacial nanobubbles (INBs)[10,11]. Many studies have indicated that formation of INBs mainly occurs on hydrophobic substrates[10-14]. In this work, we prepared heterogeneous samples by depositing mechanically exfoliated graphene flakes on mica substrates. Both graphene and mica were freshly prepared before experiments, ensuring the cleanliness of the sample surfaces and minimizing contamination (Methods). We demonstrated that AFM imaging of gas adsorption provides an alternative method to determine the local wetting properties of a surface with sub-10 nm resolution. Our data further indicated that graphene, and likely other hydrophobic surfaces and hydrophobic domains of molecules as well, effectively captures gas molecules dissolved in water, which has important implications for numerous research fields such as physics, chemistry, biology, and medicine, and may lead to new technological applications.

The graphene-coated mica samples we prepared (Methods) typically contain graphene layers of different thickness. We characterized our samples in air using optical microscopy, micro-Raman measurements, and AFM (Supplementary Fig. 1). Raman measurements of graphene layers and the mica substrate (Supplementary Fig. 1b,e) corresponded well with the results of AFM (Supplementary Fig. 1c,d). Single-layer graphene was often found at the edges of the graphene flakes. The quality of the transferred graphene was very good, as evidenced by absence of the D peak at ~1350 cm$^{-1}$ (Supplementary Fig. 1e), which is associated with disordered carbon atoms or defects.

After water injection, thin graphene flakes may detach completely from the mica substrate because the mica surface is highly hydrophilic and water enters the graphene-mica interface easily. To minimize this possibility, we kept the water injection speed below 1 ml/s. In our experiments, single-layer graphene still has a high tendency to roll into a scroll[15]. Nonetheless, occasionally we detected single-layer graphene along with thicker graphene layers on mica. Fig. 1a shows a typical example and scrolling (indicated with white arrowheads) at some edges was evident. In addition, intercalation of water layers at the mica-graphene interface produced atomically flat plateaus; the edge of a single water layer intercalation appeared very similar to a graphene atomic step (green arrowheads in Fig. 1b). Two-dimensional islands (yellow arrowheads in Fig. 1a-c) were due to the intercalation of second water layers. Similar flat plateaus with intercalation of water layers were previously reported in AFM studies of graphene-coated mica substrates in air at relatively high humidity[16].

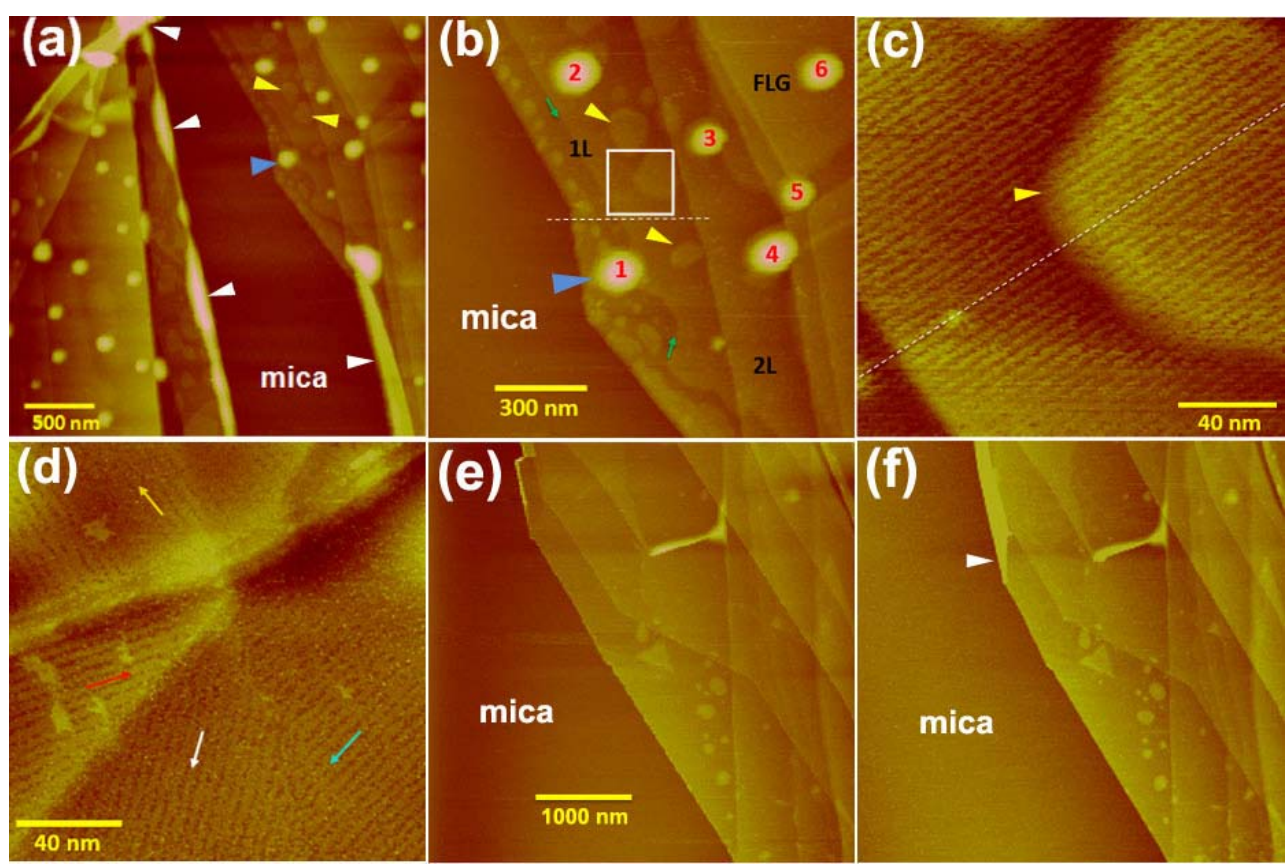

**Figure 1 Height images (AFM) of graphene sheets on mica after immersion in chilled water and in degassed water.** (**a**) Image of a graphene-coated mica sample acquired at t=40 min after chilled water (gas-supersaturated water) was deposited. Bright cap-shaped structures are present on graphene-covered regions, but not on the mica substrate, which remains dark and flat. The blue arrowhead indicates a cap-shaped structure, which serves as a marker in other images. (**b**) A higher-resolution image of the upper right-hand region in (a). The cap-shaped structures in this image were numbered and their contact angles were estimated (Supplementary Table 1). Graphene sheets with thickness of one, two, and a few layers are indicated with 1L,

2L, and FLG, respectively. A height profile along the white dashed line appears in Supplementary Figure 2a. (**c**) High-resolution image of the region inside the white box in (b). The striped pattern exhibits no phase slip (white dashed line), indicating a single stripe domain in this image. This pattern suggests that the stripe domain was on one graphene layer and that the change in step height near the lower left-hand corner was due to intercalation of a water layer under the graphene. (**d**) Height image of another graphene-covered region showing stripe domains with different row orientations (coloured arrows). These orientations are either along the zig-zag or the arm-chair directions of the graphene. (**e**) Image of another graphene-coated mica sample acquired at t=94 min after pre-degassed water was deposited. Note the absence of the cap-shaped and striped structures with degassed water. (**f**) Images of approximately the same region as in (e) at t=380 min. The features apparent after the deposition of chilled water remained absent at this late time point. Note the visualization of graphene scrolling at this very small length scale (arrowhead).

Surprisingly, bright, cap-shaped structures preferentially appeared on graphene-covered regions, while bare mica remained atomically flat and with no trace of gas adsorption (Fig. 1a,b). In addition, we observed striped patterns (row-like structures) covering almost the entire flat region of graphene outside the cap-shaped structures. Fig. 1c shows a high-resolution image inside the white box in Fig. 1b, which exhibits a striped pattern across the entire image. These data support the hypothesis that the flat plateaus were caused by intercalation of some material (probably water) between the graphene layer and mica. We confirmed that the striped pattern was a real surface structure, rather than an artefact of our data-acquisition system based on the following observations. When we acquired an image with the fast scan direction rotated 90° clockwise relative to that in Figure 1c, the stripe orientation also rotated accordingly (Supplementary Fig. 2b). In addition, row spacing, measured as 4.5±0.5 nm, did not change when we changed the scan size. Further, we observed another graphene-covered region with stripe domains with different orientations (Fig. 1d). We note that no striped pattern was evident on bare mica (data not shown).

The cap-shaped and row-like structures on graphene-coated regions (Fig. 1a-d) did not appear when pre-degassed water (Methods) was used (Fig. 1e). Even 380 min after water immersion, almost no structure was evident (Fig. 1f), but scrolling of single-layer graphene at the edges occurred over time (white arrowhead in Fig. 1f). These observations demonstrate that these structures were formed via the adsorption of air gas dissolved in water.

When we compared the maps of height (Supplementary Fig. 3a) and stiffness (Supplementary Fig. 3b) acquired with PeakForce mode, the cap-shaped nanostructures on graphene-covered regions exhibited a darker contrast in stiffness than the other regions, resembling the INBs reported elsewhere[10,11,17,18]. The force curves measured on the cap-shaped structures also exhibited characteristics that were very similar to those of INBs[17-19] (Supplementary Fig. 2c). We thus conclude that the cap-shaped structures were INBs.

The striped patterns on the graphene-covered regions (Fig. 1c,d and Supplementary Fig. 2b) resemble the row-like structures observed previously at a highly ordered pyrolytic graphite (HOPG)-water interface[17,18,20-22], which were ascribed to nitrogen- or oxygen-containing structures. However, typical domain sizes at the HOPG-water interface are on the order of tens of nm[17,18,20-22], much smaller than those observed at the graphene-water interface, which are on the order of hundreds of nm or larger. Similar striped patterns of large domains were also reported for graphene in air[23], and recently they were concluded to be responsible for friction anisotropy measured on graphene[24]. In the present study, the striped pattern on graphene-coated regions under ambient air typically appeared more than seven days after the sample was prepared. Our observation that no striped pattern occurred when the graphene-on-mica sample was immersed in degassed water (Fig. 1e,f) further indicates that the pattern forms through adsorption of nitrogen or oxygen molecules dissolved in water, and that it does not form readily in air. Recently, the striped patterns were suggested to be interfacial gas hydrates occurring at graphite-water interfaces[18], explaining the high stiffness (Supplementary Fig. 3b) and high stability of the structure.

We measured the gas-side contact angles of INBs on the graphene-covered regions in Figure 1b from the cross-section profile of each INB. The contact angles of INBs were 11~13° (Supplementary Table 1). Graphene-layer thickness did not affect the contact angle of INBs; we detected only a slight increase in the contact angle on single-layer graphene compared with few-layer graphene (Supplementary Table 1). Thus, the wetting property of single-layer graphene-coated mica is more HOPG-like than mica-like, implying that single-layer graphene already shields the majority of the interactions between water and the underlying mica substrate.

When water was rapidly heated to 45 °C before deposition (Methods), with a gas concentration of ~60% of the saturation concentration at room temperature, cap-shaped structures formed on graphene-covered regions only (Fig. 2a). Interestingly, we detected a narrow strip of single-layer graphene with a width of ~20 nm (a graphene nanoribbon) (white arrow in Fig. 2a) . Cap-shaped structures also formed on this graphene nanoribbon. Higher-resolution imaging revealed a cap-shaped structure (an INB; Fig. 2b). The INB appeared to extend laterally over the edge of the graphene nanoribbon to the mica substrate (Fig. 2b). Surprisingly, the height profiles across the INB in the directions along and perpendicular to the long axis of the graphene nanoribbon exhibited considerably higher aspects ratios than those observed on larger graphene flakes (Fig. 2c). We estimated the contact angles of INBs 1 and 2 to be 14° and 15°, respectively, which are within the typical range of INB contact angles previously reported for various hydrophobic-water interfaces (5-25°)[10,11]. In contrast, the contact angles estimated from profiles A and B (Fig. 2b) are 61° and 87°, respectively, considerably larger than those of typical INBs. This observation illustrates the intricate wetting phenomenon for nano-patterned structures and demonstrates that the graphene nanoribbon can change the local surface chemistry of the hydrophilic mica substrate. Gas adsorbates therefore appear to be much more wettable on atomically thin graphene than on mica, suggesting that the propensity of gas adsorption to a surface may be used to characterize the local surface wettability with high spatial resolution. In contrast, WCA measurements are macroscopic over a lateral dimension of 1 mm or larger. Surface defects and heterogeneities, such as step edges, impurities, and adsorbates, are almost unavoidable on interfaces of the macroscopic scale, and affect the WCA measurement values[8].

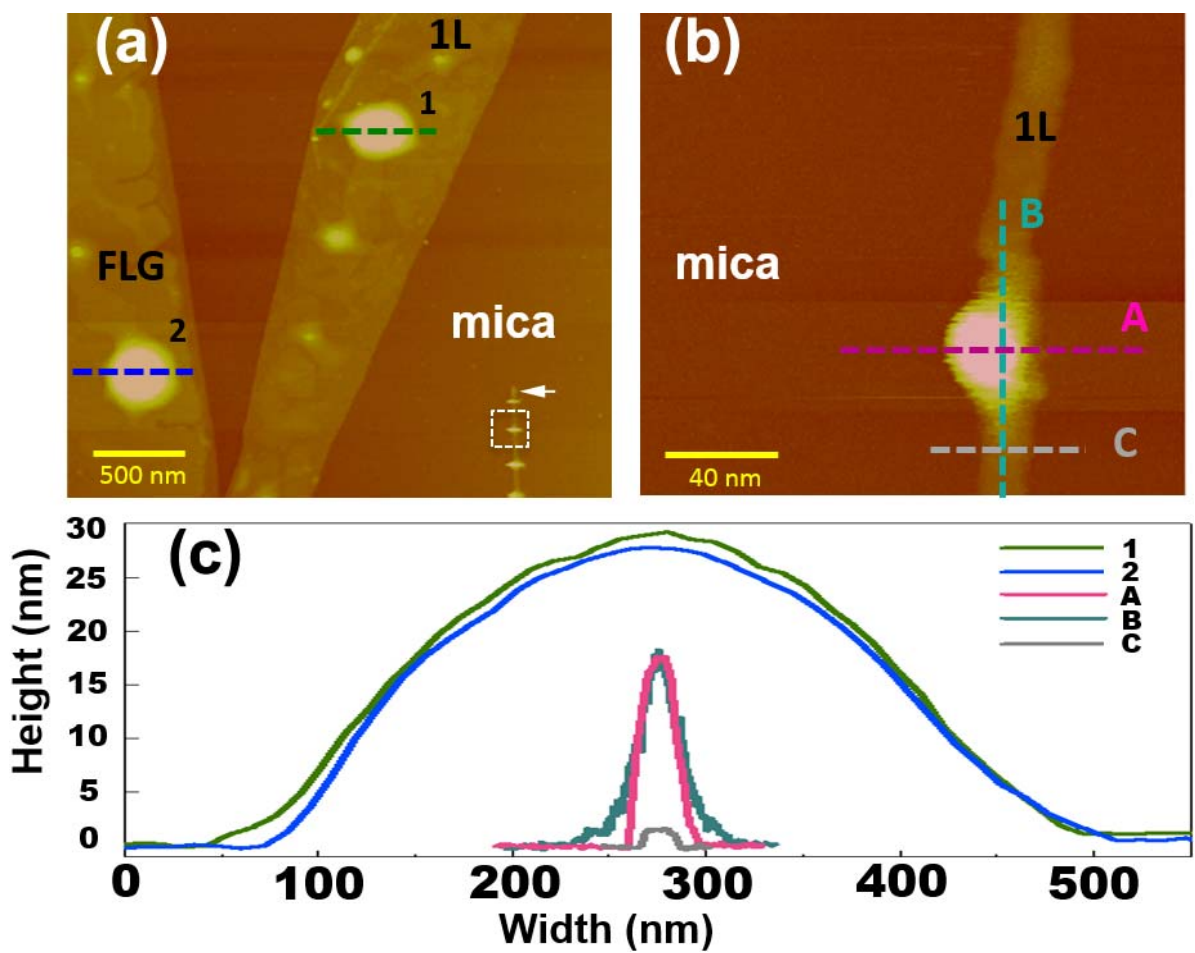

**Figure 2 PeakForce measurements of a graphene-coated mica after rapidly heated water was deposited.** (**a**) Topographic image (AFM) acquired at t=90 min after water deposition. Two INBs on graphene are numbered and their height profiles along the dashed line are shown in (c). A white arrow indicates a graphene nanoribbon. 1L, one-layer graphene sheet; FLG, few-layer graphene sheet. (**b**) Higher-resolution image of the white boxed region in (a). (**c**) Height profiles across cap-shaped INBs along trace lines 1, 2, A, B, and C in (a) and (b).

We conducted five experiments similar to the one shown in Figure 2 using rapidly heated water; INBs formed on graphene-covered regions in all cases (data not shown). In most experiments and applications in water without deliberate control over gas concentration, the air concentration is close to the saturation concentration. Thus, gas adsorption should occur in typical situations with graphene in contact with water. Therefore, it is essential for theoretical modelling to consider the presence of gas molecules at graphene-water interfaces for comparison with experiments.

Why do dissolved gas molecules prefer to be adsorbed at the graphene-water interface over the mica-water interface? It has been proposed that water forms hydration layers on highly hydrophilic surfaces such as mica, presenting multiple energy barriers to approaching objects[25]. Kinetically, it is difficult for dissolved gas molecules to penetrate through the hydration layers to reach the mica substrate. Graphene is more hydrophobic than mica; thus there may be no significant hydration-layer barriers at graphene-water interfaces. In addition, the attractive hydrophobic interactions between dissolved gas molecules (in particular, gas clusters) and a hydrophobic surface may facilitate gas adsorption. Thermodynamically, hydrophobic surfaces may provide low-chemical-potential sites at which gas molecules dissolved in water can be adsorbed[17].

The preferential adsorption of dissolved gas molecules on hydrophobic surfaces may have profound and widespread implications for many phenomena in physics, chemistry, and biology. It has been reported that gas bubbles tend to form on hydrophobic surfaces versus hydrophilic surfaces in water[26]. Preferential gas adsorption may hence constitute an initial step for subsequent bubble formation. The inner surfaces of blood vessels are hydrophobic[27], and were suggested to serves as sites for gas micronuclei during bubble formation on decompression[26,28]. The current study highlights the possibility of adsorption and accumulation of nitrogen and/or oxygen molecules on the inner surfaces of blood vessels; the effects of this adsorption on biological functions should be investigated in future studies. In addition, preferential gas adsorption may be responsible for the drag reduction for water flow over a hydrophobic surface (boundary slip)[29], and many electrode surfaces, such as HOPG, Au, and Pt, are hydrophobic, suggesting that the formation of gas-containing structures may affect electrochemical reactions. Adsorption of gas, particularly the formation of INBs, may also affect the adsorption and bonding of molecules or particles to a hydrophobic surface in aqueous solutions. This behaviour may have beneficial effects, such as the prevention of surface fouling[30], or undesirable effects, such as non-uniform immobilization of a molecular layer on gold surfaces for cantilever-based biosensors[19]. Moreover, it may be interesting to study whether preferential gas adsorption is related to the breathing of aquatic organisms for efficient capture of oxygen molecules dissolved in water. It is known that heme proteins, which contain iron porphyrins, play a crucial role in oxygen storage and transport (for example by myoglobin and hemoglobin). Porphyrin molecules are hydrophobic and may greatly increase the cross section for the capture of oxygen molecules in blood. Finally, on a global level, gas segregation may affect the structural stablility and self-assembly behaviors of organic/biological molecules, because most of these molecules contain hydrophobic components.

## Methods

### Preparation of graphene-coated mica samples

The topmost layers of two mica plates (TED PELLA 50-12, 12 mm × 12 mm) and an HOPG sample (lateral sizes of 12 mm × 12 mm, ZYB; Momentive) were first peeled off with scotch tape to expose clean surfaces and immediately transferred to a glove box. Subsequent processes did not involve tape, polymers, or chemicals, ensuring that the prepared samples were clean. The cleaved mica plates were placed on a hot plate (200 °C) to prevent water condensation on mica surfaces. A small piece of a thin graphite flake was carefully removed from the HOPG surface with tweezers and placed on the cleaved surface of a mica plate. The glove box was then sealed and purged by pumping out the air from one side of the box and simultaneously injecting high-purity $N_2$ gas from the other side in order to reduce the humidity inside the glove box. The humidity and temperature were measured with a hygrometer inside the glove box. When the humidity reached <20%, the freshly cleaved side of the other mica plate was pressed strongly against the graphite flake by hand for ~1 min (Supplementary Fig. 4a). Then, the glove box was opened, and the top mica plate was removed. The graphite flake was detached from the bottom mica substrate with tweezers. Usually, several graphene flakes with different numbers of graphene layers remained adhered to the mica substrate. The samples were first sorted using optical microscopy and then imaged with AFM. The number of graphene layers was determined via Raman spectroscopy and AFM.

### Water preparation

All water was purified using a Milli-Q system (Millipore Corp.) at a resistivity of 18.2 MΩ·cm. Slightly air-supersaturated water was prepared by storing purified Milli-Q water with air in a sealed 50-ml conical centrifuge tube at 4 °C in a refrigerator for several days. Before AFM, the chilled water was taken out of the refrigerator and either (i) deposited on a graphene-coated mica sample or (ii) heated to 45 °C in a ~100°C hot bath before water deposition (Supplementary Fig. 4b). For water deposition, the water was extracted using a disposable syringe (5 ml) and injected slowly into the commercial AFM liquid cell onto a graphene-coated mica sample. To estimate the air concentration in water, the oxygen concentration of water was measured using a dissolved oxygen tracer (Lamotte 1761M). For the chilled water (~4 °C), the oxygen concentration was measured as 11~12 mg/l. When the chilled water was heated to 45 °C, the oxygen concentration reduced to ~5 mg/l, or ~60% of the saturation oxygen concentration at 25 °C (8.3 mg/l). Gas bubbles were evident at the inner wall of the centrifuge tube, explaining the decreased oxygen concentration in this rapidly heated water. The water volume in the AFM liquid cell was only 60 μl, and the water temperature was expected to stabilize near room temperature within 10 s after injection into the liquid cell. Unless otherwise specified, all experiments were carried out at room temperature.

For a few experiments, degassed water was deposited on the graphene-coated mica sample. Degassed water was prepared by storing the freshly purified deionized water in a desiccator, which was then pumped to 0.1~0.2 atm for several minutes and sealed for >15 h until immediately before AFM. A syringe was used to extract the degassed water and inject it into a liquid cell for AFM. The oxygen concentration was 10-20% of the saturation oxygen concentration at 25 °C.

**AFM**

AFM was performed in PeakForce mode using a Bruker AXS Multimode NanoScope V at room temperature (23-25 °C). In this mode, the sample is oscillated in the vertical direction with amplitude of tens to hundreds of nanometers and at a frequency of 2 kHz (Supplementary Fig. 4c). Vertical piezo movement results in cycles of approaching and retracting traces that lead to force-distance curves in which the tip makes intermittent contact with the sample surface. Topography information is obtained from the height correction performed by the feedback loop to keep a constant "peak" of force, while the slope of the contact region determines the stiffness of the sample at each pixel. Si cantilevers (OMCL-AC240TS from Olympus) with a spring constant of ~0.7-3.8 N/m were used, and the nominal tip radius was ~10 nm. Before AFM, a sample was placed in a commercial fluid cell. Typically the sample was first scanned in air (without water injection). After the sample was immersed in water, further AFM measurements were conducted. The peak force for acquiring the images in this work was set at 200 pN, which is the minimum for stable imaging of entire regions. The oscillation amplitude was 20 nm.

**Micro-Raman spectroscopy**

A micro-Raman system (WITec alpha 300) with a laser with an excitation wavelength of 532 nm was used to characterize graphene-coated mica samples. The laser spot size was ~1 μm. Raman spectra were measured at specific sites. The system also allows 2D mapping of a specific peak. Raman measurements were conducted in air.

Acknowledgements

We thank the National Science Council (NSC99-2112-M-001-029-MY3 and NSC102-2112-M-001-024-MY3), the Ministry of Science and Technology (MOST 103-2627-M-001 -011 and MOST 104-2627-M-001-005) of the Republic of China, and Academia Sinica for supporting this study.

# Supplementary Information

## High-resolution characterization of preferential gas adsorption at the graphene-water interface

Hsien-Chen Ko[1], Wei-Hao Hsu[1,2], Chih-Wen Yang[1], Chung-Kai Fang[1], Yi-Hsien Lu[1], Ing-Shouh Hwang[1,2]*

Index of the Supplementary Information:

1. Supplementary Figures

2. Supplementary Table

3. Supplementary References

# Supplementary Figures

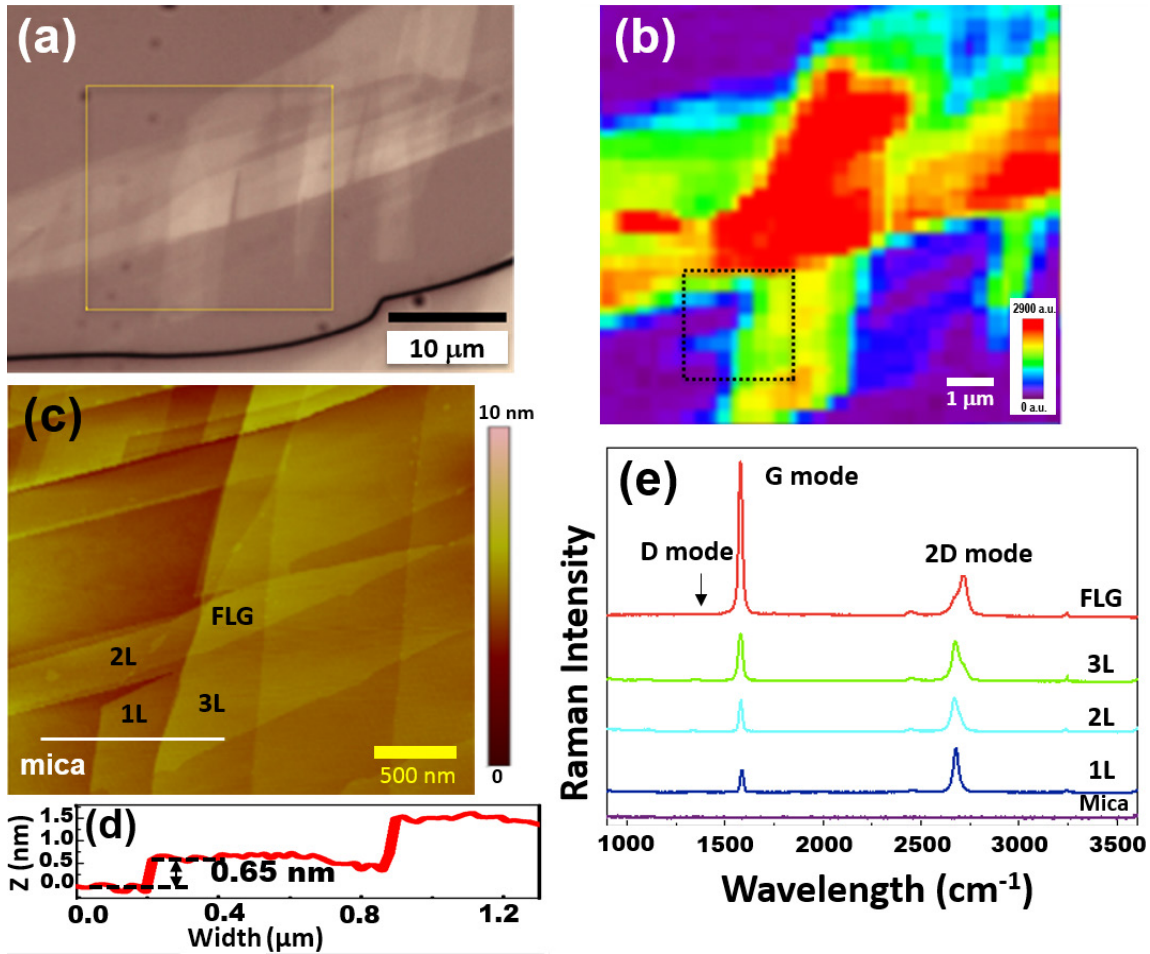

**Supplementary Figure 1 Characterization of a graphene-coated mica sample in air with optical microscopy, Raman microscopy, and AFM.** (**a**) Bright-field optical microscopy of a sample; the graphene-covered area has a lateral dimension of tens of microns. (**b**) Raman map plotted according to the intensity of the G band of graphene inside the area outlined by a yellow box in (a). Pure mica regions (purple) did not exhibit any G peak. Single-layer graphene regions are shown in deep blue. (**c**) AFM of the area outlined by a black box in (b). 1L, single-layer graphene; 2L, double-layer graphene; 3L, triple-layer graphene; FLG, few-layer graphene. (**d**) Height profile along the white line in (c). The height profile indicates that single-layer graphene is 0.65 nm higher than the mica substrate, which is within the range of 0.3-1.5 nm previously reported[1]. This range of apparent step height was attributed to the chemical and electrostatic differences between graphene and mica[2]. (**e**) Raman spectra measured at regions of bare mica and

differing numbers of graphene layers. Colours are as in (b). The ratio of the 2D peak intensity (at ~2700 cm$^{-1}$) to the G peak intensity (at ~1580 cm$^{-1}$) decreased as the number of graphene layers increased, consistent with previous reports[3]. The D mode at ~1350 cm$^{-1}$ is marked. It is associated with disordered carbon atoms or defects[4] and absence of this peak indicates good quality of the graphene sample.

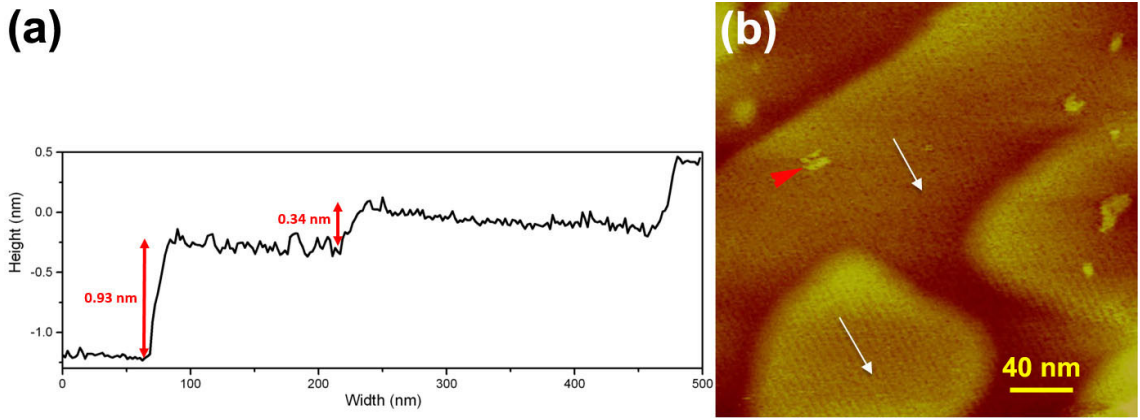

**Supplementary Figure 2 AFM of a graphene-coated mica sample in water.** (**a**) Height profile along the white dashed line in Figure 1b of the main text. (**b**) Height image acquired with the fast scan direction rotated 90° clockwise relative to that in Figure 1c of the main text. The white arrows indicate the row orientation of the stripe pattern. The red arrowhead indicates a bright structure forming on the stripe pattern. Several other bright structures were evident on the striped pattern; their lateral size was small and they changed positions over time. These characteristics were similar to our previous observations of the second-layer ordered structure on the row-like pattern at the HOPG-water interface[5].

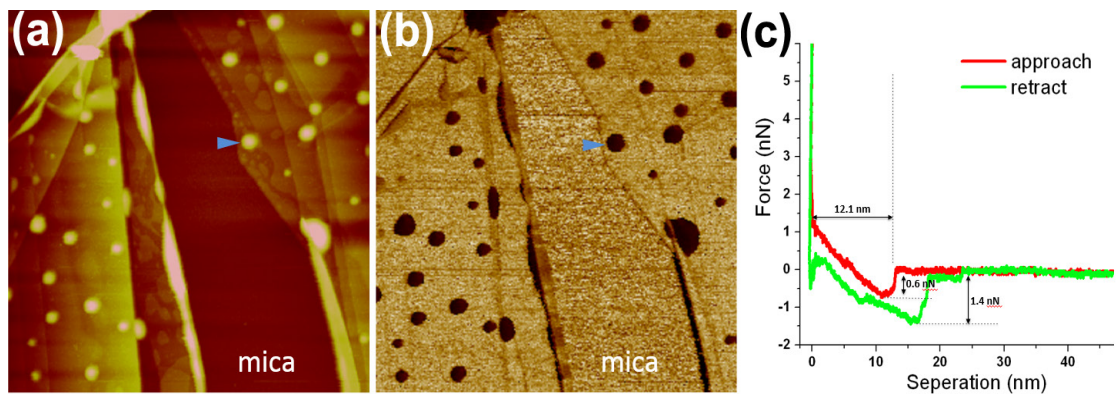

**Supplementary Figure 3 PeakForce measurements of graphene sheets on mica after deposition of chilled water.** (**a**) Height image. (**b**) The corresponding stiffness map. INBs exhibited dark contrast, indicating that they are softer than other regions of the surface. (**c**) Force curve measured on the cap-shaped structure marked with a blue arrow in (a).

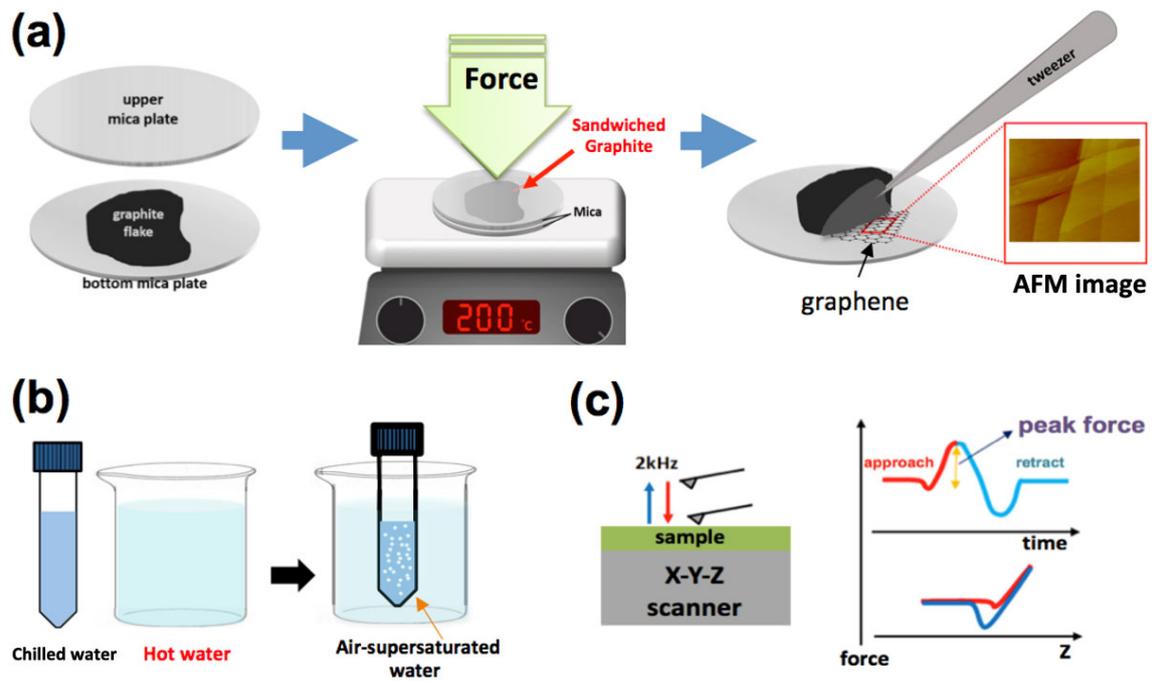

**Supplementary Figure 4 Sample and water preparation for AFM.** (**a**) Schematic of the preparation of graphene-coated mica samples. (**b**) Illustration of water preparation. (**c**) Illustrations of peak force mode. See Methods for details.

# Supplementary Table

| INB | Height (nm) | Width (nm) | Contact angle (°) |
|-----|-------------|------------|-------------------|
| 1 | 8.9 | 152 | 13 |
| 2 | 8.8 | 160 | 13 |
| 3 | 6.8 | 129 | 12 |
| 4 | 9.6 | 176 | 13 |
| 5 | 6.7 | 129 | 12 |
| 6 | 7.5 | 153 | 11 |

**Supplementary Table 1**. Apparent height, width, and calculated contact angle of INBs on graphene with different numbers of layers.

# Supplementary References